\documentclass[pra,preprint,superscriptaddress,showpacs,tightenlines]{revtex4}

\usepackage{graphics}           
\usepackage{bm}                 
\usepackage{amsmath}            
\usepackage{amssymb}
\usepackage{verbatim}           
\usepackage{amsthm}             
\theoremstyle{plain}            

\numberwithin{equation}{section}

\def\bra#1{{\langle#1|}}
\def\ket#1{{|#1\rangle}}
\def\bracket#1#2{{\langle#1|#2\rangle}}

\def\expect#1{{\langle#1\rangle}}
\def\e{{\rm e}}
\def\proj{{\hat{\cal P}}}
\def\tr{{\rm Tr}}
\def\H{{\hat H}}

\def\Lop{{\cal L}}
\def\Ahat{{\hat A}}
\def\Adag{{\hat A}^\dagger}
\def\Ehat{{\hat E}}

\def\Shat{{\hat S}}
\def\Sdag{{\hat S}^\dagger}
\def\U{{\hat U}}
\def\Udag{{\hat U}^\dagger}
\def\Zhat{{\hat Z}}

\def\Op{{\hat O}}
\def\id{{\hat I}}
\def\Pr{{\proj_{\rm R}}}
\def\Pl{{\proj_{\rm L}}}
\def\x{{\hat x}}

\begin{document}

\title{Quantum random walks with decoherent coins}

\author{Todd A. Brun}
\email{tbrun@ias.edu}
\affiliation{Institute for Advanced Study, Einstein Drive, Princeton, 
              NJ 08540}

\author{H. A. Carteret}
\email{hcartere@cacr.math.uwaterloo.ca}
\affiliation{Department of Combinatorics and Optimization, 
             University of Waterloo, Waterloo, Ontario, N2L 3G1, Canada}

\author{Andris Ambainis}
\email{ambainis@ias.edu}
\affiliation{Institute for Advanced Study, Einstein Drive, Princeton, 
              NJ 08540}

\date{2002}

\begin{abstract}
The quantum random walk has been much studied recently, largely due
to its highly nonclassical behavior.  In this paper, we study one
possible route to classical behavior for the discrete
quantum walk on the line:  the presence of decoherence in
the quantum ``coin'' which drives the walk.  We find exact analytical
expressions for the time dependence of the first two moments of
position, and show that in the long-time limit the variance grows
linearly with time, unlike the unitary walk.  We compare this to the
results of direct numerical simulation, and see how the form of the
position distribution changes from the unitary to the usual classical
result as we increase the strength of the decoherence.
\end{abstract}


\maketitle

\section{Introduction}

In the classical discrete random walk, a particle is located at one of
a set of definite positions (such as the set of integers on the line).
In response to a random event---for example, the flipping
of a coin---the particle moves either right or left.  This
process is iterated, and the motion of the particle is analyzed
statistically.  These systems provide good models for diffusion
and other stochastic processes.

Considerable work has been done recently on quantum random walks,
which are unitary (and hence reversible) systems
designed as analogues to the usual classical case.  There are two
general approaches to the problem:  {\it continuous}
\cite{FarhiGutmann98,Childs01,Childs02} and {\it discrete}
\cite{Aharonov93,Meyer96,NayaknV,Aharonov00,Ambainis01,Meyer01,Moore01,
QW:higherdim,Kempe02,Konno02,Konno02b,Konno02c,Travaglione02,Du02,Yamasaki02,
Sanders02,Duer02,Bach02,Kendon02} unitary walks.
This paper is exclusively concerned with the discrete walk.
In this discrete case, we introduce an extra ``coin'' degree of
freedom (usually a single quantum bit) into the system.
Just as in the classical random walk, the outcome of a ``coin flip''
determines which way the particle moves; but in the quantum case,
both the ``flip'' of the coin and the conditional motion of the particle
are unitary transformations.  Different possible classical paths
can interfere with each other.

In this paper we look at quantum walks on the infinite line.  The 
particle is initially at position $x=0$ and is free to travel off to 
infinity in either direction.  We will look at both
the probability distribution in $p(x,t) = \bra{x}\rho_t\ket{x}$, and at
the long-time behavior of the moments $\expect{\x}$ and
$\expect{\x^2}-\expect{\x}^2$ as functions of $t$.

For a classical random walk, $p(x,t)$ has the form of a
binomial distribution, with a width which spreads like $\sqrt{t}$;
the variance ${\bar{x^2}}-{\bar x}^2$ grows linearly with time.
The variance in the quantum walk, by contrast, grows {\it quadratically}
with time; and the distribution $p(x,t)$ has a complicated, oscillatory
form.  Both of these are effects of interference between the
possible paths of the particle.

It should be possible to recover the classical behavior as some kind
of limit of the quantum system.  There are two obvious ways to regain
the classical result.  If the quantum ``coin'' is measured at every step,
then the record of the measurement outcomes singles out a particular
classical path.  By averaging over all possible measurement records,
one recovers the usual classical behavior \cite{QW:higherdim}.

Alternatively, rather than re-using the same coin every time, one could
replace it with a {\it new} quantum coin for each flip.  After a time
$t$ one would have accumulated $t$ coins, all of them entangled with
the position of the particle.  By measuring them, one could reconstruct
an unique classical path; averaging over the outcomes would once again
produce the classical result.

These two approaches, which are equivalent in the classical limit,
give two different routes from quantum to classical \cite{Letter}.
We might increase the number of coins used to generate the walk, cycling
among $M$ different coins, in the limit using a new coin at each step.
Or we might {\it weakly} measure the coin after each step, reaching the
classical limit with strong, projective measurements.  This is equivalent
to having a coin which is subject to {\it decoherence}.

In another paper \cite{Multicoin} we have considered the quantum random walk
with multiple coins.  In this case, the quantum behavior remains
qualitatively unchanged until we reach the limit of a new coin for
each step, at which point classical behavior is recovered.

In this paper, we consider the quantum random walk with a single coin
subject to decoherence.  We will see that in this case, the behavior
is qualitatively quite different from the unitary quantum random walk.
The usual classical solution is recovered in the limit where the coin
decoheres completely every step; but even with weaker decoherence,
the variance of the position distribution grows linearly with time,
rather than quadratically (as in the unitary case).

In section II we present an analytical result for the moments of the
decoherent walk, and compare them to the results from direct numerical
simulations.  In section III, we see how the probability distribution
$p(x,t)$ changes as we introduce decoherence.  Finally, in section IV
we summarize our results and state conclusions.

\section{Moments of the decoherent walk}

\subsection{The unitary walk on the line}

Let us now consider a fairly general quantum random walk on the line.
The particle degree of freedom has a basis of position eigenstates
$\{\ket{x}\}$ where $x$ can be any integer.  The position operator
is $\x$, and $\x\ket{x} = x\ket{x}$.  We will assume that the
particle begins the walk at the origin, in state $\ket0$.  The walk
is driven by a separate
``coin'' degree of freedom:  a $D$-dimensional
system with an initial state $\ket{\Phi_0}$.  Let $\Pr,\Pl$ be two
orthogonal projectors on the Hilbert space of the ``coin,'' such that
$\Pr + \Pl = \id$.  These represent the two possible outcomes of
the coin flip:  Heads or Tails, Right or Left.
We also define a unitary transformation $\U$
which ``flips'' the coin by rotating a coin showing heads or tails
into a superposition of the two.  One step of the quantum random walk is
given by the unitary operator
\begin{equation}
\Ehat \equiv \left(\Shat \otimes \Pr + \Sdag \otimes \Pl\right)
  \left( \id \otimes \U \right) \;,
\end{equation}
where $\Shat,\Sdag$ are unitary shift operators on the particle position:
\begin{equation}
\Shat\ket{x} = \ket{x+1} \;, \ \ 
\Sdag\ket{x} = \ket{x-1} \;.
\end{equation}
The full initial state of the system (particle and ``coin'') is
\begin{equation}
\ket{\Psi_0} = \ket0 \otimes \ket{\Phi_0} \;.
\end{equation}

We can identify the eigenvectors $\ket{k}$ of $\Shat,\Sdag$,
\begin{equation}
\ket{k} = \sum_x \e^{ikx}\ket{x} \;,
\end{equation}
with eigenvalues
\begin{eqnarray}
\Shat\ket{k} &=& \e^{-ik}\ket{k} \;, \nonumber\\
\Sdag\ket{k} &=& \e^{+ik}\ket{k} \;.
\end{eqnarray}
The inverse transformation is
\begin{equation}
\ket{x} = \int_{-\pi}^{\pi} \frac{dk}{2\pi} \e^{-ikx}\ket{k} \;.
\end{equation}
In particular, the initial state of the particle is
\begin{equation}
\ket{0} = \int_{-\pi}^{\pi} \frac{dk}{2\pi} \ket{k} \;.
\end{equation}
These state vectors $\ket{k}$ are not renormalizable, but if used with
caution they greatly simplify the calculations.  
In the $k$ basis, the evolution operator becomes
\begin{eqnarray}
\Ehat\left( \ket{k} \otimes \ket\Phi \right)
  &=& \ket{k} \otimes \left( \e^{-ik} \Pr + \e^{ik} \Pl\right) \U
  \ket\Phi \;, \nonumber\\
  &\equiv& \ket{k} \otimes \U_k \ket\Phi \;,
\end{eqnarray}
where $\U_k$ is also a unitary operator.

The usual case considered in the literature has taken the coin
to be a simple two-level system, and the ``flip'' operator $\U$
to be the usual Hadamard transformation $\H$:
\begin{eqnarray}
\H \ket{R} &=& \frac{1}{\sqrt{2}} \left( \ket{R} + \ket{L} \right) \;, 
  \nonumber\\
\H \ket{L} &=& \frac{1}{\sqrt{2}} \left( \ket{R} - \ket{L} \right) \;.
\end{eqnarray}
The projectors are $\Pr = \ket{R}\bra{R}$, $\Pl = \ket{L}\bra{L}$.
The walk on the line in this case has been exactly solved by
Nayak and Vishwanath \cite{NayaknV}.

For the present, we will continue without assuming a particular
form for $\U$, $\Pr$ or $\Pl$.  Later we will specialize to make
comparison to numerical simulations.

\subsection{Decoherence}

We now generalize to allow for decoherence.  Suppose
that before each unitary ``flip'' of the coin, a completely positive
map is performed on the coin (note, NOT on both the coin and the
particle).  This map is given by a set of operators $\{\Ahat_n\}$
on the coin degree of freedom which satisfy
\begin{equation}
\sum_n \Adag_n\Ahat_n = \id \;.
\end{equation}
A density operator $\chi$ for the coin degree of freedom is transformed
\begin{equation}
\chi \rightarrow \chi' = \sum_n \Ahat_n \chi \Adag_n \;.
\end{equation}
If we apply this to a general density operator for the joint
particle/coin system
\begin{equation}
\rho = \int \frac{dk}{2\pi}\int \frac{dk'}{2\pi} \ket{k}\bra{k'} \otimes
  \chi_{kk'} \;,
\end{equation}
then after one step the state becomes
\begin{equation}
\rho \rightarrow \rho'
  = \int \frac{dk}{2\pi}\int \frac{dk'}{2\pi} \ket{k}\bra{k'} \otimes
  \sum_n \U_k \Ahat_n \chi_{kk'} \Adag_n \Udag_{k'} \;.
\end{equation}
The initial state is
\begin{equation}
\rho_0 = \int \frac{dk}{2\pi}\int \frac{dk'}{2\pi} \ket{k}\bra{k'} \otimes
  \ket{\Phi_0}\bra{\Phi_0} \;.
\end{equation}
Let the quantum random walk proceed for $t$ steps.  Then the state
evolves to
\begin{equation}
\rho_t = \int \frac{dk}{2\pi}\int \frac{dk'}{2\pi} \ket{k}\bra{k'} \otimes
  \sum_{n_1,\ldots,n_t} \U_k \Ahat_{n_t} \cdots \U_k \Ahat_{n_1}
  \ket{\Phi_0}\bra{\Phi_0}
  \Adag_{n_1} \Udag_{k'} \cdots \Adag_{n_t} \Udag_{k'}  \;.
\end{equation}
We define the {\it superoperator} $\Lop_{kk'}$ on the coin degree of freedom
\begin{equation}
\int \frac{dk}{2\pi}\int \frac{dk'}{2\pi} \ket{k}\bra{k'} \otimes
  \sum_n \U_k \Ahat_n \chi_{kk'} \Adag_n \Udag_{k'} \equiv
  \int \frac{dk}{2\pi}\int \frac{dk'}{2\pi} \ket{k}\bra{k'}
  \otimes \Lop_{kk'} \chi_{kk'} \;.
\end{equation}
In terms of the superoperator,
\begin{equation}
\rho_t = 
  \int \frac{dk}{2\pi}\int \frac{dk'}{2\pi} \ket{k}\bra{k'} \otimes
  \Lop_{kk'}^t  \ket{\Phi_0}\bra{\Phi_0} \;.
\end{equation}
Note that for $k=k'$ this superoperator preserves the trace.
This implies that
\begin{equation}
\tr\left\{ \Lop_{kk}^n \Op \right\} =
  \tr\left\{ \Op \right\} \;,
\label{traceidentity}
\end{equation}
for any operator $\Op$.  This identity will prove useful later.

The probability to reach a point $x$ at time $t$ is
\begin{eqnarray}
p(x,t) &=&  \tr\left\{ \ket{x}\bra{x} \rho_t \right\} =
  \bra{x}\rho_t\ket{x} \nonumber\\
&=& \frac{1}{(2\pi)^2} \int dk \int dk' \bracket{k}{x}\bracket{x}{k'}
  \tr\left\{ \Lop_{kk'}^t \ket{\Phi_0}\bra{\Phi_0} \right\}
  \nonumber\\
&=& \frac{1}{(2\pi)^2} \int dk \int dk' \e^{-ix(k-k')}
  \tr\left\{ \Lop_{kk'}^t \ket{\Phi_0}\bra{\Phi_0} \right\} \;.
\label{probdistribution}
\end{eqnarray}

\subsection{Moments of position}

Eq.~(\ref{probdistribution}) for $p(x,t)$ will be difficult to evaluate,
in general.  However, we can get considerably further by restricting
our interest to the {\it moments} of this distribution.
\begin{eqnarray}
\expect{\x^m}_t &=& \sum_x x^m p(x,t) \nonumber\\ 
&=& \frac{1}{(2\pi)^2} \sum_x x^m \int dk \int dk' \e^{-ix(k-k')}
  \tr\left\{ \Lop_{kk'}^t \ket{\Phi_0}\bra{\Phi_0} \right\} \;.
\end{eqnarray}
We can then invert the order of operations and do the $x$ sum first.
This sum can be exactly carried out in terms of derivatives of the
delta function:
\begin{equation}
\frac{1}{2\pi} \sum_x x^m \e^{-ix(k-k')}
  = (-i)^m \delta^{(m)}(k-k') \;.
\end{equation}
Inserting this result back into our expression for $\expect{\x^m}_t$ yields
\begin{equation}
\expect{\x^m}_t =
  \frac{(-i)^m}{2\pi} \int dk \int dk' \delta^{(m)}(k-k')
  \tr\left\{ \Lop_{kk'}^t \ket{\Phi_0}\bra{\Phi_0} \right\} \;.
\end{equation}
We can then integrate this by parts.

In carrying out this integration by parts, we will need
\begin{eqnarray}
\frac{d}{dk} \tr\left\{ \Lop_{kk'} \Op \right\}
  &=& \tr\left\{ (d\Lop_{kk'}/dk) \Op \right\} \nonumber\\
&=& \sum_n \tr\left\{ (d\U_k/dk) \Ahat_n \Op
  \Adag_n \Udag_{k'} \right\} \;,
\label{lopderiv}
\end{eqnarray}
where
\begin{eqnarray}
\frac{d\U_k}{dk} &=&
  - i (\proj_0 - \proj_1) \U_k \equiv - i \Zhat \U_k \nonumber\\
\frac{d\Udag_k}{dk} &=&
  i \Udag_k (\proj_0 - \proj_1) \equiv i \Udag_k \Zhat \nonumber\\
\Zhat &=& \Pr - \Pl = \id - 2 \Pl \;.
\end{eqnarray}
Substituting this back into (\ref{lopderiv}) we get
\begin{eqnarray}
\frac{d}{dk} \tr\left\{ \Lop_{kk'} \Op \right\}
  &=& - i \tr\left\{ \Zhat \Lop_{kk'} \Op \right\} \;, \nonumber\\
&=& - i \tr\left\{ (\Lop_{kk'} \Op) \Zhat \right\} \;, \nonumber\\
&=& - \frac{d}{dk'} \tr\left\{ \Lop_{kk'} \Op \right\} \;.
\label{lopderiv2}
\end{eqnarray}

Making use of (\ref{traceidentity}) and (\ref{lopderiv2}), when we
carry out the integration by parts for the first moment we get
\begin{equation}
\expect{\x}_t = - \frac{1}{2\pi} \sum_{j=1}^t \int dk
  \tr\left\{ \Zhat \Lop_{kk}^j \ket{\Phi_0}\bra{\Phi_0} \right\} \;.
\label{firstmoment1}
\end{equation}
We can simplify our notation slightly by defining
\begin{equation}
\Lop_{k} \equiv \Lop_{kk} \;.
\end{equation}

We can carry out a similar integration by parts to get the second moment:
\begin{eqnarray}
\expect{\x^2}_t &=& - \frac{1}{2\pi} \int dk \Biggl[
 \sum_{j=1}^t \sum_{j'=1}^j 
  \tr\left\{ \Zhat \Lop_k^{j-j'} ( \Zhat \Lop_k^{j'}
  \ket{\Phi_0}\bra{\Phi_0} ) \right\} \nonumber\\
&& + \sum_{j=1}^t \sum_{j'=1}^{j-1} 
  \tr\left\{ \Zhat \Lop_k^{j-j'} ( ( \Lop_k^{j'}
  \ket{\Phi_0}\bra{\Phi_0} ) \Zhat ) \right\} \Biggr] \;.
\label{secondmoment1}
\end{eqnarray}
Note that this form is rather similar to that for a correlation function.

For the unitary walk \cite{Letter,Multicoin}, we were able to
separate the expressions for
the moments into oscillatory and nonoscillatory terms, and consider
only the nonoscillatory terms for the long-time limit.  In the present
case, this will not work; because of the presence of decoherence, the
oscillations are damped.  However, the long-time limit still simplifies
in another way.  If we think of $\Lop_k$ as a linear transformation,
all its eigenvalues must clearly obey $|\lambda| \le 1$ in order for
it to be completely positive.  In finding the long-time limit,
we need pick out only those components of the expressions (\ref{firstmoment1})
and (\ref{secondmoment1}) which do not die away at large $t$.

\subsection{The Hadamard walk with decoherence}

To get any further than this, we need to specialize to a particular
model.  Let's choose the standard two-dimensional ``coin'' and the
Hadamard walk as described above.  This makes our operator $\U_k$
\begin{equation}
\U_k = \frac{1}{\sqrt{2}}
\begin{pmatrix} \e^{-ik}&\e^{-ik}\\
  \e^{ik}&-\e^{ik}
\end{pmatrix} \;.
\label{onecoinUk}
\end{equation}
We need also to pick a particular form for the decoherence.  I will
choose the decoherence produced by the three operators
\begin{eqnarray}
\Ahat_0 = \sqrt{p} \ket{R}\bra{R} \;, \nonumber\\
\Ahat_1 = \sqrt{p} \ket{L}\bra{L} \;, \nonumber\\
\Ahat_2 = \sqrt{1-p} \id \;.
\label{decoherence}
\end{eqnarray}
This resembles a coin which has a probability $p$ per step of being
measured; however, this is equivalent to a large class of other decoherence
models, such as {\it pure dephasing}
\begin{eqnarray}
\Ahat'_0 = \frac{1}{\sqrt2} \left( \e^{i\theta} \ket{R}\bra{R} +
  \e^{-i\theta} \ket{L}\bra{L} \right) \;, \nonumber\\
\Ahat'_1 = \frac{1}{\sqrt2} \left( \e^{-i\theta} \ket{R}\bra{R} +
  \e^{i\theta} \ket{L}\bra{L} \right) \;,
\label{dephasing}
\end{eqnarray}
or {\it weak measurement}
\begin{eqnarray}
\Ahat''_0 = \sqrt{q} \ket{R}\bra{R} +
  \sqrt{1-q}\ket{L}\bra{L} \;, \nonumber\\
\Ahat''_1 = \sqrt{1-q} \ket{R}\bra{R} +
  \sqrt{q} \ket{L}\bra{L} \;.
\label{weak}
\end{eqnarray}
All three of these decoherence processes represent the same completely
positive map on the density matrix; one can transform from one to
the other by invoking the relationships between the parameters
\begin{equation}
1-p = \cos 2\theta = 2 \sqrt{q(1-q)} \;.
\end{equation}
We will generally use the first in our derivations, for simplicity;
but the second is more convenient for numerical calculations.  Note
that this equivalence applies to the {\it average} behavior over all
outcomes; for a single sequence of outcomes, the conditional dynamics
of these three decoherence (or weak measurement) models can look quite
different.

In the absence of decoherence, the Hadamard walk exhibits a
linear drift in $\expect{\x}$ and a quadratic growth in the
variance $\expect{\x^2}-\expect{\x}^2$.  We plot these moments
in figure 1.

\begin{figure*}[t]
\includegraphics{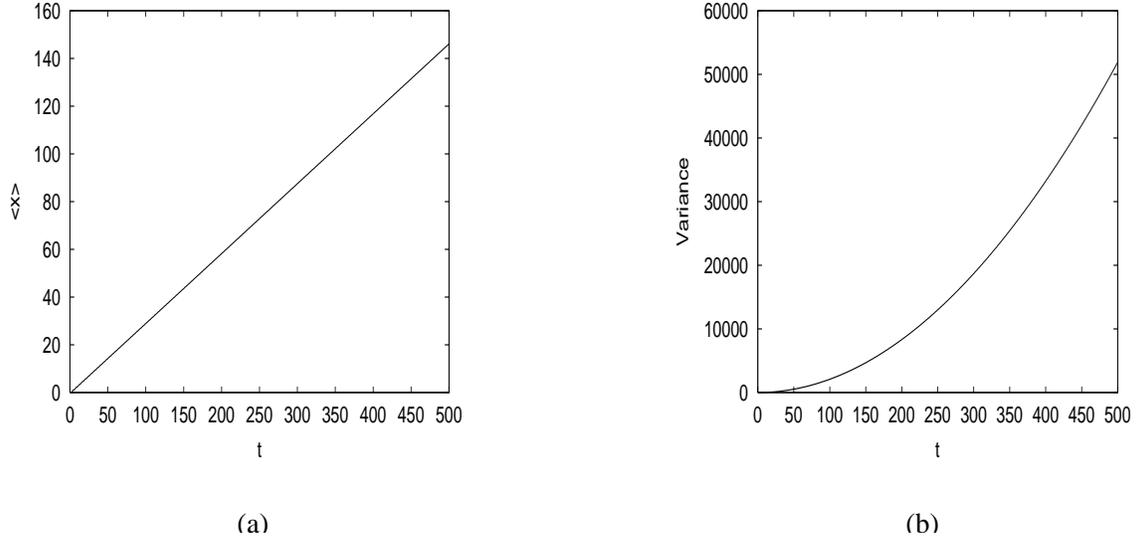}
\caption{\label{fig1}  The (a) expectation $\expect\x$, and (b) variance
$\expect{\x^2}-\expect{\x}^2$, for the Hadamard walk on the line without
decoherence.  The coin begins in state $\ket{R}$.  Note that the
first moment exhibits a linear drift with time, which seems to reflect
a peculiar ``memory'' for the initial state; a system starting with the
coin in state $\ket{L}$ would drift symmetrically in the other direction.
The variance grows quadratically in time, in contrast to the linear
growth in the classical random walk.  Note that if viewed at a finer
scale, both of these curves would exhibit oscillations about the simple
power-law behavior.}
\end{figure*}

Because $\Lop_k$ is linear, we can represent it as a matrix acting
on the space of two-by-two operators.  A convenient representation is to
write
\begin{equation}
\Op = r_0 \id + r_1 \sigma_1 + r_2 \sigma_2 + r_3 \sigma_3 \;,
\end{equation}
where $\sigma_{1,2,3} = \sigma_{x,y,z}$ are the usual Pauli matrices.
We can then represent $\Op$ by a column vector
\begin{equation}
\Op \equiv
\begin{pmatrix} r_0 \\ r_1 \\ r_2 \\ r_3
\end{pmatrix} \;.
\end{equation}
The action of $\Lop_k$ on $\Op$ is then given by
\begin{equation}
\Lop_k \Op \equiv
\begin{pmatrix}
  1 & 0 & 0 & 0 \\
  0 & 0 & -(1-p)\sin 2k & \cos 2k \\
  0 & 0 & -(1-p)\cos 2k & - \sin 2k \\
  0 & 1-p & 0 & 0
\end{pmatrix}
\begin{pmatrix} r_0 \\ r_1 \\ r_2 \\ r_3
\end{pmatrix} \;.
\label{lopmatrix}
\end{equation}
Note that since $r_0 = \tr\{\Op\}$, it is unaffected by $\Lop_k$, which
is trace-preserving.  So the only nontrivial dynamics result from
the three-by-three submatrix
\begin{equation}
M_k \equiv
\begin{pmatrix}
  0 & -(1-p)\sin 2k & \cos 2k \\
  0 & -(1-p)\cos 2k & - \sin 2k \\
  1-p & 0 & 0
\end{pmatrix} \;.
\label{mkmatrix}
\end{equation}

We also need to know the effects of left-multiplying and right-multiplying
by $\Zhat$.  These are given by the two matrices
\begin{equation}
Z_L \equiv
\begin{pmatrix}
  0 & 0 & 0 & 1 \\
  0 & 0 & i & 0 \\
  0 & -i & 0 & 0 \\
  1 & 0 & 0 & 0
\end{pmatrix} \;, \ \ 
Z_R \equiv
\begin{pmatrix}
  0 & 0 & 0 & 1 \\
  0 & 0 & -i & 0 \\
  0 & i & 0 & 0 \\
  1 & 0 & 0 & 0
\end{pmatrix} \;.
\end{equation}
Finally, taking the trace picks out the 0 component of the column
vector and drops the rest.

Let us take these expressions and apply them to equation
(\ref{firstmoment1}) for the first moment.  In the integrand,
the initial density matrix for the coin is multiplied $j$ times
by $\Lop_k$, then left-multiplied by $\Zhat$, and finally the
trace is taken.  Given the above expression for $Z_L$, we see that
this is the same as multiplying the three-vector $(r_1,r_2,r_3)$
$j$ times by $M_k$ and then keeping only the $r_3$ component of
the result.  This gives us the new expression:
\begin{eqnarray}
\expect{\x}_t &=& - \frac{1}{2\pi} \int dk
\begin{pmatrix}
  0 & 0 & 1
\end{pmatrix}
  \left[ \sum_{j=1}^t M_k^j \right]
\begin{pmatrix}
  r_1 \\ r_2 \\ r_3
\end{pmatrix} \nonumber\\
&=& - \frac{1}{2\pi} \int dk
\begin{pmatrix}
  0 & 0 & 1
\end{pmatrix}
  \left[ (1-M_k)^{-1} (M_k - M_k^{t+1}) \right]
\begin{pmatrix}
  r_1 \\ r_2 \\ r_3
\end{pmatrix} \;.
\label{firstmoment2}
\end{eqnarray}
The eigenvalues of $M_k$ are complicated, but fortunately we don't
need to evaluate them.  All we need to know is that all of them
obey $0 < |\lambda| < 1$ (where both these inequalities are strict).
In the long time limit, therefore, $M_k^{t+1} \rightarrow 0$, and
the moment becomes approximately
\begin{equation}
\expect{\x}_t \approx - \frac{1}{2\pi} \int dk
\begin{pmatrix}
  0 & 0 & 1
\end{pmatrix}
  \left[ (1-M_k)^{-1} M_k \right]
\begin{pmatrix}
  r_1 \\ r_2 \\ r_3
\end{pmatrix} \;.
\label{firstmoment2b}
\end{equation}
Note that all $t$ dependence has vanished.  Therefore, in the long
time limit, the first moment tends to a constant.

As it happens, the matrix $1-M_k$ is exactly invertible:
\begin{equation}
(1-M_k)^{-1} = \frac{1}{p(2-p)}
\begin{pmatrix}
  1 + (1-p)\cos 2k & - (1-p)\sin 2k & (1-p)+\cos 2k \\
  - (1-p)\sin 2k & 1 - (1-p)\cos 2k & - \sin 2k \\
  (1-p)(1+(1-p)\cos 2k) & - (1-p)^2\sin 2k & 1 + (1-p)\cos 2k 
\end{pmatrix} \;.
\label{mkinvmatrix}
\end{equation}
Inserting this and (\ref{mkmatrix}) into (\ref{firstmoment2b}) yields
\begin{eqnarray}
\expect{\x}_t &\approx& - \frac{1}{2\pi} \int dk
\frac{1-p}{p(2-p)} \left[ (1-p) r_3 + r_1
  + (r_3 + (1-p) r_1) \cos 2k
  - (1-p) r_2 \sin 2k \right] \nonumber\\
&=& \frac{1-p}{p(2-p)} \left[ (1-p) r_3 + r_1 \right] \nonumber\\
&=& \frac{1-p}{p(2-p)} \left[ (1-p) (|\alpha|^2 - |\beta|^2)
  + (\alpha^*\beta + \alpha\beta^*) \right] \;,
\label{firstmoment3}
\end{eqnarray}
where the initial state of the coin is
$\ket{\Phi_0} = \alpha\ket{R} + \beta\ket{L}$.

In figure 2, we compare this result to the results of direct numerical
simulation for the initial state $\ket{\Phi_0} = \ket{R}$.  As we can
see, the first moment does tend to drift towards a constant value
asymptotically.  What variation there is is a result of statistical
error in the Monte Carlo simulation rather than a poor fit.

\begin{figure}[t]
\includegraphics{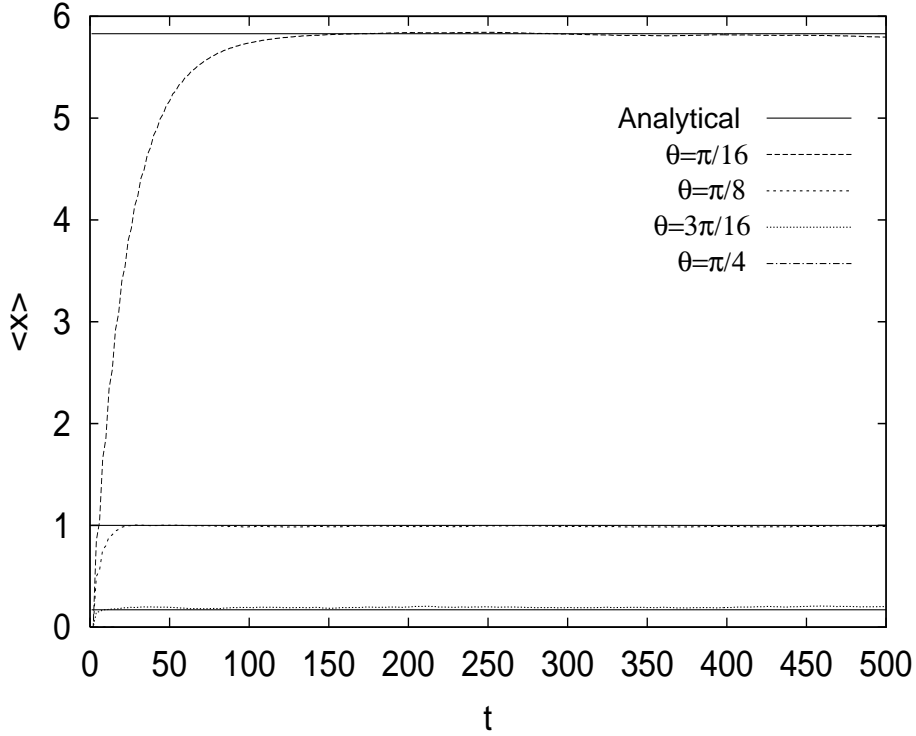}
\caption{\label{fig2}  $\expect\x$ vs. $t$ for the Hadamard walk on
the line with decoherence, for $\theta=\pi/16,\pi/8,3\pi/16,\pi/4$.
For all cases the coin began in the initial
state $\ket{R}$.  Note that $\expect\x$ goes asymptotically to a
constant value at long times which matches our analytical estimate;
this drift goes to zero with increasing decoherence, vanishing
at $\theta=\pi/4$ (i.e., $p=1$).  Note that the irregularities in the
broken curves reflect statistical errors in the Monte Carlo calculations.}
\end{figure}

The second moment is a somewhat more complicated calculation.  Let us
rewrite the equation (\ref{secondmoment1}) in terms of our four-by-four
matrices:
\begin{eqnarray}
\expect{\x^2}_t &=& - \frac{1}{2\pi} \int dk \Biggl[
 \sum_{j=1}^t \sum_{j'=1}^j 
  \tr\left\{ \Zhat \Lop_k^{j-j'} ( \Zhat \Lop_k^{j'}
  \ket{\Phi_0}\bra{\Phi_0} ) \right\} \nonumber\\
&& + \sum_{j=1}^t \sum_{j'=1}^{j-1} 
  \tr\left\{ \Zhat \Lop_k^{j-j'} ( ( \Lop_k^{j'}
  \ket{\Phi_0}\bra{\Phi_0} ) \Zhat ) \right\} \Biggr] \nonumber\\
&=& t - \frac{1}{2\pi} \int dk
\begin{pmatrix} 1 & 0 & 0 & 0
\end{pmatrix}
\left[ Z_L \sum_{j=1}^t \sum_{j'=1}^{j-1} \Lop_k^{j-j'}
  (Z_L + Z_R) \Lop_k^{j'} \right]
\begin{pmatrix} 1 \\ r_1 \\ r_2 \\ r_3
\end{pmatrix} \nonumber\\
&=& t - \frac{1}{2\pi} \int dk
\begin{pmatrix} 0 & 0 & 0 & 1
\end{pmatrix}
\left[ \sum_{j=1}^t \sum_{j'=1}^{j-1} \Lop_k^{j-j'}
  (Z_L + Z_R) \Lop_k^{j'} \right]
\begin{pmatrix} 1 \\ r_1 \\ r_2 \\ r_3
\end{pmatrix} \;.
\end{eqnarray}
The initial $t$ term comes from the $j=j'$ components of
(\ref{secondmoment1}), where the two factors of $\Zhat$ cancel out.

Recall the block-diagonal form (\ref{lopmatrix}) of $\Lop_k$, and
note that
\begin{equation}
(Z_L + Z_R) =
\begin{pmatrix}
  0 & 0 & 0 & 2 \\
  0 & 0 & 0 & 0 \\
  0 & 0 & 0 & 0 \\
  2 & 0 & 0 & 0
\end{pmatrix}
\end{equation}
is extremely sparse.  In this case, it makes sense to separate the
0 and $1,2,3$ components of the expression:
\begin{eqnarray}
\expect{\x^2}_t &=& t - \sum_{j=1}^t \sum_{j'=1}^{j-1} \frac{1}{2\pi} \int dk
\begin{pmatrix} 0 & 0 & 0 & 1
\end{pmatrix}
\Biggl[ \Lop_k^{j-j'}
  (Z_L + Z_R) \Lop_k^{j'}
\begin{pmatrix} 1 \\ 0 \\ 0 \\ 0
\end{pmatrix} \nonumber\\
&& + \Lop_k^{j-j'}
  (Z_L + Z_R) \Lop_k^{j'}
\begin{pmatrix} 0 \\ r_1 \\ r_2 \\ r_3
\end{pmatrix} \Biggr] \;.
\end{eqnarray}

We can drop the second term, because it will be cancelled in the
inner product with
$\begin{pmatrix} 0 & 0 & 0 & 1
\end{pmatrix}$.  This means that $\expect{\x^2}_t$ has {\it no dependence}
on the initial state!  The remaining term becomes
\begin{eqnarray}
\expect{\x^2}_t &=& t - \sum_{j=1}^t \sum_{j'=1}^{j-1} \frac{1}{2\pi} \int dk
\begin{pmatrix} 0 & 0 & 0 & 1
\end{pmatrix}
\Lop_k^{j-j'}
\begin{pmatrix} 0 \\ 0 \\ 0 \\ 2
\end{pmatrix} \nonumber\\
&=& t - \frac{1}{2\pi} \int dk
\begin{pmatrix}  0 & 0 & 1
\end{pmatrix}
\left[ \sum_{j=1}^t \sum_{j'=1}^{j-1} M_k^{j-j'} \right]
\begin{pmatrix} 0 \\ 0 \\ 2
\end{pmatrix} \nonumber\\
&=& t - \frac{1}{2\pi} \int dk
\begin{pmatrix}  0 & 0 & 1
\end{pmatrix} \nonumber\\
&& \times \left[ t - (1-M_k)^{-1} M_k
  + (1-M_k)^{-1} M_k^t  \right] (1-M_k)^{-1} M_k
\begin{pmatrix} 0 \\ 0 \\ 2
\end{pmatrix} \;.
\label{secondmoment_exact}
\end{eqnarray}

In the long-time limit, we can drop the $M_k^t$ terms as
negligible.  We can thereby substitute the exact expressions (\ref{mkmatrix})
and (\ref{mkinvmatrix}) for the remaining matrices and simplify:
\begin{eqnarray}
\expect{\x^2}_t &=& t + \frac{1}{2\pi} \int dk
  \frac{1-p}{p(2-p)} \Biggl[ 2 t (1-p+\cos 2k) \nonumber\\
&& - \frac{1}{p(2-p)} \left( 4 (2-2p+p^2) \cos 2k
  + (1-p)(7+\cos 4k) \right) \Biggr] \nonumber\\
&=& t \left(1 + \frac{2(1-p)^2}{p(2-p)} \right)
  - \frac{7(1-p)^2}{p^2(2-p)^2} \;.
\label{secondmoment2}
\end{eqnarray}
Thus, in the case of the decoherent coin the variance will grow
linearly with time at long times, just as in the classical case, though
the rate of growth will be greater than for the classical random walk.
Note that as $p\rightarrow1$, $\expect{\x^2}_t \rightarrow t$, which is
the classical result.

Because of the dependence of the first moment on the initial condition,
the variance will have some residual dependence on the initial condition;
but since this is only a constant it is quite unimportant.
Upon comparison to the numerical results (see figure 3),
we see that this analytical
expression for the variance matches extremely well.

\begin{figure}[t]
\includegraphics{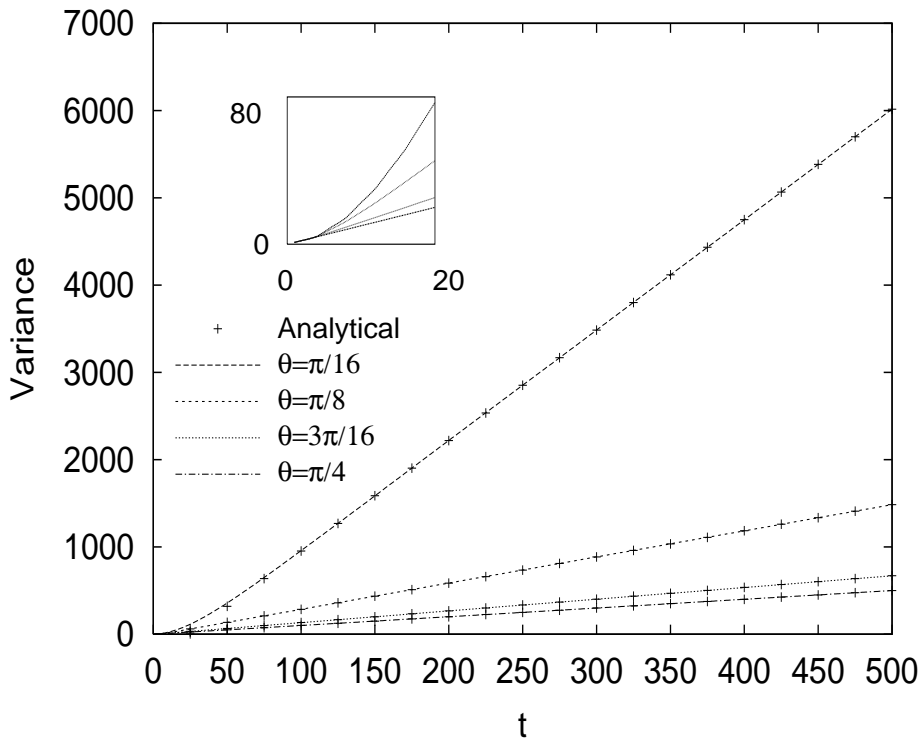}
\caption{\label{fig3}  $\expect{\x^2}-\expect{\x}^2$ vs.~$t$ for the
quantum random walk with decoherence, for $\theta=\pi/16,\pi/8,3\pi/16,\pi/4$.
For all cases the coin began in the initial
state $\ket{R}$.  Note that the variance goes asymptotically to a
linear growth at long times which matches our analytical estimate;
the rate of growth goes to one with increasing decoherence, matching
the classical case at $\theta=\pi/4$ (i.e., $p=1$).  The inset shows
the time-dependence of the variance at short times, where it still
exhibits quadratic growth.}
\end{figure}

Since this approximation holds in the long-time limit, it is reasonable
to ask how long is a ``long time.''  For weak decoherence ($p \ll 1$),
the evolution exhibits the same behavior as the unitary walk
for short times, with the second moment growing quadratically in time,
and only switches over to linear growth past some critical timescale.
This changeover is visible in figure 3.

We take the long-time approximation by neglecting the $M_k^t$ term
in (\ref{secondmoment_exact}).  If we examine $M_k^2$, we see that
it can be written as $(1-p)$ multiplying a matrix whose determinant is
less than one.  Therefore, if we want this term to be negligible, then
\begin{equation}
||M_k^t|| < (1-p)^{t/2} < \epsilon
\end{equation}
for some positive $\epsilon \ll 1$.  This implies that
\begin{eqnarray}
p > (2/t) \log (1/\epsilon) \equiv c/t
\end{eqnarray}
which implies in turn that the second moment of position will grow
linearly at $t > c/p$, for a large enough constant $c$.  For
$t < c/p$ there will be a transition over to quadratic time dependence.
For small $p$, equation (\ref{secondmoment2}) becomes
\begin{equation}
\expect{\x^2}_t =
 =  t \left(1 + 1/p \right) - 7/4p^2 \;.
\end{equation}

\section{The position distribution}

As mentioned before, Eq.~(\ref{probdistribution}) for $p(x,t)$ is
difficult to evaluate analytically.  It is straightforward, however,
to solve this system numerically.  Rather than solve for the density
matrix $\rho$, we instead have used quantum trajectory techniques
\cite{Trajectory} to do a quantum Monte Carlo simulation, averaging over
many runs to find both the distribution itself and its moments.

For the purposes of these simulations, it proved more convenient to
use the form (\ref{dephasing}) for the decoherence process.  We have
therefore labeled the figures with the appropriate values of the
dephasing parameter $\theta$ instead of $p$.  In performing the
simulations, after each step of the quantum walk we randomly applied
either operator $\sqrt{2}\Ahat'_0$ or $\sqrt{2}\Ahat'_1$ from
eq.~(\ref{dephasing}) to the state, choosing them with equal probability.
Because $\sqrt{2}\Ahat'_{0,1}$ are unitary operators, it was unnecessary
to renormalize the state.  We then found the mean probability
distributions $p(x)$ by averaging over many runs;
each distribution plotted in this paper represents an average
over $10000$ runs of the Monte Carlo program.
In figure 4 we see how the probability distribution changes as we go
to the classical limit of complete decoherence at every step
($\theta = \pi/4$).

\begin{figure*}[t]
\includegraphics{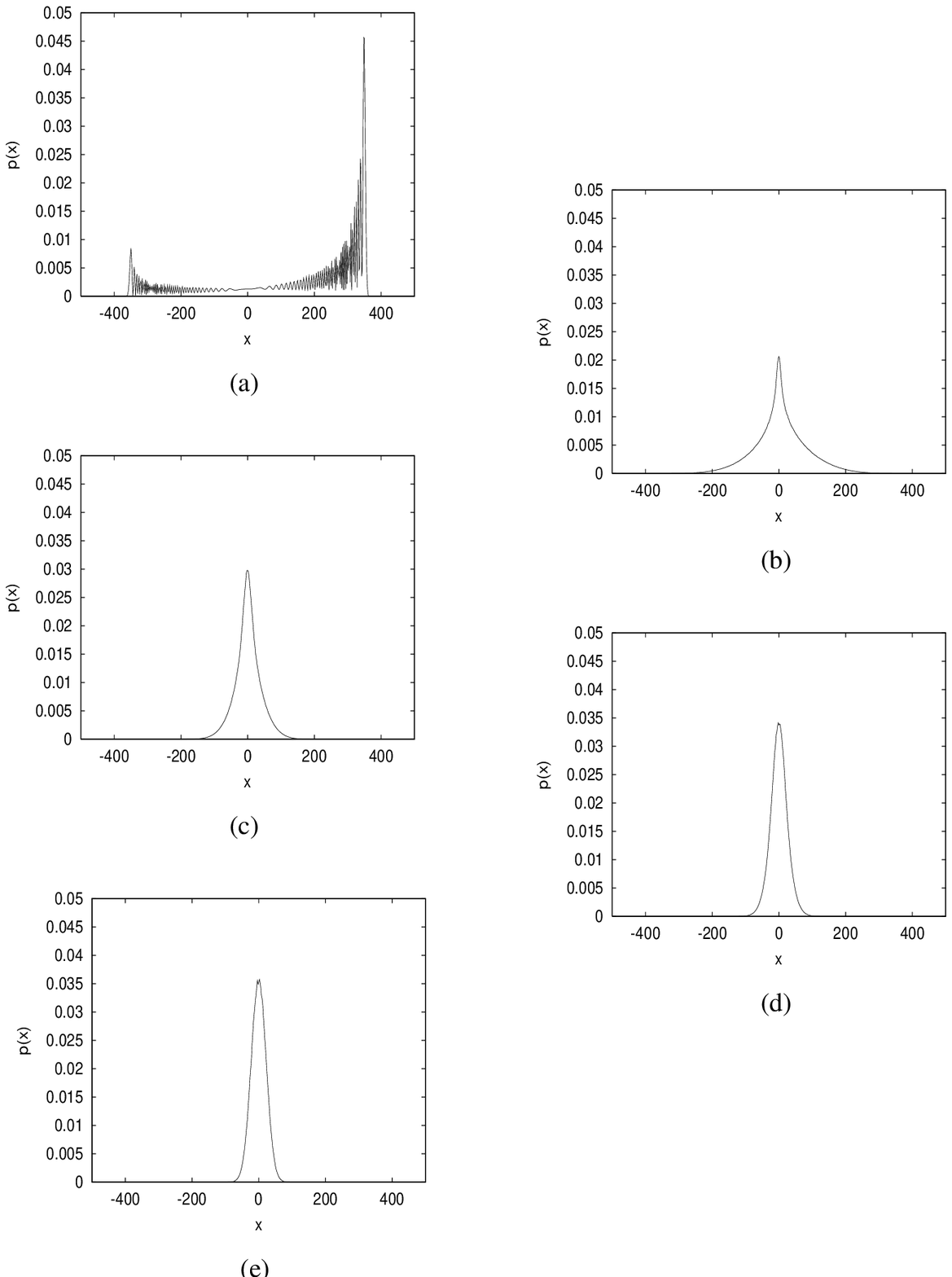}
\caption{\label{fig4}  The probability distributions $p(x,t)$
at $t=500$ for the quantum random walk with decoherence:
(a) $\theta=0$ (unitary case), (b) $\theta=\pi/16$, (c) $\theta=\pi/8$
(d) $\theta=3\pi/16$, (e) $\theta=\pi/4$ (classical case).  A central
peak appears when decoherence is included, which becomes
increasingly dominant; however, the peak is broadened compared to the
classical case, with nontrivial tails which disappear only in the
classical limit, when $p(x,t)$ goes over to the binomial distribution.}
\end{figure*}

As we see from the figure, the presence of decoherence quickly wipes
out the most conspicuous signs of interference:  the peaks at
$\pm t/\sqrt{2}$ and the oscillations in the distribution.  A new
peak, centered at $t=0$, makes its appearance, though for weak decoherence
it does not have the usual binomial form.  However, some residual
effects of interference persist up until just short of the classical
limit.  In particular, the distribution is broadened, with long ``tails,''
and displaced somewhat from the central position.
This is reflected in our long-time solution for the variance
(\ref{secondmoment2}), which is higher in the quantum
case than the classical except in the limit $p\rightarrow 1$.

In the unitary case, interference effects cause the evolution to
``remember'' the starting state in a non-intuitive way.  As
we saw in figure 1, if we start the coin in state $\ket{R}$, the
system retains a linear drift to the right for all time.  If we had
instead started in the state $\ket{L}$, the system would have drifted
symmetrically to the left.  This ``memory'' also makes the position
distribution asymmetric, as we see in fig.~4a.

The presence of decoherence eliminates this effect at long times:
the system ``forgets'' the initial conditions after a while.
This effect will still persist at short times, however, and
produce a tendency for the particle to move repeatedly in the
same direction.  No doubt this tendency produces the broadening
in the position distribution and the significant tails.

\section{Conclusions}

We have examined a possible route from quantum to classical
for the quantum random walk:  letting the quantum coin undergo
decoherence with time.  Using the long-time behavior of the
position moments as a qualitative marker of classicality, we
see that even very weak decoherence changes the growth of the
variance from a quadratic to a linear function of time.  Since
the usual classical random walk has a linear growth of the variance,
in a particular sense we can claim that the decoherent walk is
indeed classical.

This situation is quite different from the quantum random walk with
multiple coins \cite{Multicoin}; for that system, quadratic growth
remained except in the limit of a new coin for every step.  One
might reasonably claim that the multicoin system remains ``quantum''
even in the limit of very large numbers of coins, while the
decoherent system remains ``classical'' even in the limit of very
weak noise.

In spite of this, some effects of interference are important even in
the presence of decoherence.  In particular, the variance grows more
rapidly in the quantum than the classical case.  This may reflect
a tendency for the particle to be ``biased'' in one direction for
short times, while interference effects enable it to ``remember''
its starting state.  This is consistent with the position distributions,
which are broader than the classical binomial distribution, and have
nontrivial tails.

In the long time limit, the time dependence of the moments becomes
tractable because the evolution superoperator has only one eigenvalue
of modulus 1.  All other components decay away exponentially with
time.  It is interesting to speculate whether a decoherence process
with several modulus 1 eigenvalues might continue to exhibit
quantum behavior.  This is probably possible, at least for
higher-dimensional coins.

We should emphasize that the system we have studied in this paper
has decoherence only in the coin degree of freedom; any effects of
decoherence on the particle are indirect.  It might be arguably
natural to include decoherence in the particle position as well.
Such systems have been studied numerically, and exhibit interesting
effects of their own \cite{Duer02,Kendon02,Kendon02b}.

\begin{acknowledgments}

We would like to thank Bob Griffiths, Lane Hughston, Viv Kendon,
Michele Mosca and Bruce Richmond for useful conversations.
TAB acknowledges financial support from
the Martin A.~and Helen Chooljian Membership in Natural Sciences,
and DOE Grant No.~DE-FG02-90ER40542.  AA was supported by NSF grant
CCR-9987845, and by the State of New Jersey.  HAC was supported by
MITACS, The Fields Institute, and NSERC CRO project ``Quantum
Information and Algorithms.''

\end{acknowledgments}

\bibliographystyle{apsrev}
\bibliography{qwalks}


\end{document}